\documentclass[%
 reprint,
superscriptaddress,
 amsmath,amssymb,
 aps,
]{revtex4-1}

\usepackage{graphicx}
\usepackage{dcolumn}
\usepackage{bm}
\usepackage{times}
\usepackage[usenames,dvipsnames]{xcolor}

\begin{document}


\title{Geometric frustration on the trillium lattice in a magnetic metal--organic framework} 

\author{Johnathan M. Bulled}
\affiliation{Inorganic Chemistry Laboratory, University of Oxford, South Parks Road, Oxford OX1 3QR, U.K.}%
\author{Joseph A. M. Paddison}
\affiliation{Materials Science and Technology Division, Oak Ridge National Laboratory, Oak Ridge, TN 37831, U.S.A.}%
\affiliation{Churchill College, University of Cambridge, Storey’s Way, Cambridge CB3 0DS, U.K.}
\author{Andrew Wildes}
\affiliation{Institut Laue-Langevin, BP156, 71 Avenue des Martyrs, 38000, Grenoble, France}%
\author{Elsa Lhotel}
\affiliation{Institut N{\'e}el, 25 Avenue des Martyrs, 38042 Grenoble, France}%
\author{Simon J. Cassidy}
\affiliation{Inorganic Chemistry Laboratory, University of Oxford, South Parks Road, Oxford OX1 3QR, U.K.}%
\author{Breog{\'a}n Pato-Dold{\'a}n}
\affiliation{Department of Chemistry, University of Bergen, P.O. Box 7803, N-5020 Bergen, Norway}%
\author{L. Claudia G{\'o}mez-Aguirre}
\affiliation{Department of Fundamental Chemistry and CICA, Faculty of Sciences University of A Coruña, 15071 A Coruña, Spain}%
\author{Paul J. Saines}
\affiliation{School of Physical Sciences, University of Kent, Canterbury CT2 7NH, U.K.}%
\author{Andrew L. Goodwin}
\affiliation{Inorganic Chemistry Laboratory, University of Oxford, South Parks Road, Oxford OX1 3QR, U.K.}%

\date{\today}

\begin{abstract}
In the dense metal-organic framework Na[Mn(HCOO)$_3$], Mn$^{2+}$ ions ($S=\frac{5}{2}$) occupy the nodes of a `trillium' net. We show that the system is strongly magnetically frustrated: the N\'eel transition is suppressed well below the characteristic magnetic interaction strength; short-range magnetic order persists far above the N\'eel temperature; and the magnetic susceptibility exhibits a pseudo-plateau at $\frac{1}{3}$-saturation magnetisation. A simple model of nearest-neighbour Heisenberg antiferromagnetic and dipolar interactions accounts quantitatively for all observations, including an unusual 2-$\mathbf k$ magnetic ground-state. We show that the relative strength of dipolar interactions is crucial to selecting this particular ground-state. Geometric frustration within the classical spin liquid regime gives rise to a large magnetocaloric response at low applied fields that is degraded in powder samples as a consequence of the anisotropy of dipolar interactions.
\end{abstract}

\maketitle

Geometrically-frustrated magnets are of fundamental interest because their macroscopic ground state degeneracies give rise to a number of exotic effects \cite{Moessner_2006,Savary_2016,Han_2012,Starykh_2015}. The suppression of magnetic ordering to temperatures far below the dominant magnetic interaction energy scale means that small perturbations can have a profound effect on both the magnetic ground state and its excitations \cite{Moessner_2006}. For obvious reasons, a common motif in the interaction network of frustrated materials is the presence of odd-membered rings---especially triangles. Yet despite the large number of relevant topologies, much of the field has focused on the relatively small set of structure types that are common amongst ceramic materials; \emph{e.g.}, the pyrochlore, triangular, kagome, and face-centred cubic nets. The study of frustration on less common lattices may therefore allow for the discovery of novel magnetic phases and their corresponding physics \cite{Paddison_2015,Gao_2007,Kikuchi_2015}.

The chemistry of metal--organic frameworks (MOFs) allows rational design of network structures that can be difficult to realise in conventional ceramics \cite{Zhou_2012}. MOFs are comprised of inorganic `nodes' and organic `linkers'; by choosing the connectivity and geometry of each component, the chemist has remarkable control over the topology of the resulting network structure \cite{Yaghi_2003}. Likewise, the particular combination of metal and ligand employed then governs the relevant magnetic degrees of freedom, and the strength and anisotropy of their interactions \cite{Saines_2018,Thoraninsdottir_2020}.

\begin{figure}[b]
	\includegraphics{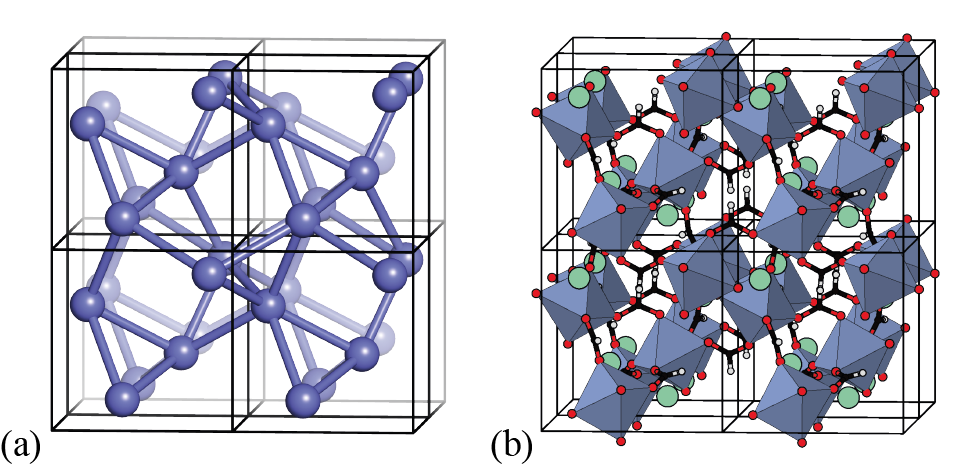}
	\caption{\label{fig1} (a) The trillium lattice, shown here as a $2\times2\times2$ supercell of the corresponding cubic unit cell \cite{Hopkinson_2005}. Each site has six neighbours, arranged at the vertices of a three-bladed propellor (\emph{cf}. the $\beta$-Mn net \cite{OKeeffe_1977}). (b) Representation of the crystal structure of Na[Mn(HCOO)$_3$]; Na, C, H and O atoms shown as green, black, white and red spheres, respectively, and Mn coordination environments shown as filled polyhedra \cite{ParedesGarcia_2009}. The Mn atoms decorate a trillium lattice.}
\end{figure}

The trillium lattice is an obvious target for applying MOF chemistry to frustrated magnetism: it is a chiral network of corner-sharing triangles that is intrinsically predisposed to geometric frustration [Fig.~\ref{fig1}(a)] \cite{Hopkinson_2005}. Remarkably few magnets are known to adopt this lattice, but those that do---mostly intermetallics---exhibit a wide range of interesting physics. In MnSi, for example, magnetic skyrmions arise from competition between ferromagnetism and the Dzyalloshinskii-Moriya interaction \cite{Muhlbauer_2009,Hopkinson_2007};  EuPtSi and EuPtGe are strongly-correlated spin liquids \cite{Franco_2017}; and CeIrSi is an Ising trillium spin-ice candidate \cite{Redpath_2010,Kneidinger_2019}. Theory tells us that even the simple nearest-neighbour Heisenberg antiferromagnet (nnHAF) on this lattice should support a classical spin-liquid (CSL) phase over a wide temperature range \cite{Hopkinson_2005}, with 120$^\circ$ helical order (``Y'' phase; $\mathbf k_{\rm ord}= [\frac{1}{3},0,0]$) in the frustrated magnetic ground state \cite{Isakov_2008}. To best of our knowledge, there has not been any experimental realisation of this model system.

It was in this context that we developed an interest in sodium manganese(II) formate, Na[Mn(HCOO)$_3$]: a dense MOF in which magnetic Mn$^{2+}$ ($S=\frac{5}{2}$) ions decorate a trillium lattice [Fig.~\ref{fig1}(b)] \cite{ParedesGarcia_2009}. The system crystallises in the chiral cubic space group $P2_13$ and has a nearest neighbour Mn--Mn distance of $\sim$5.6\,\AA. Formate ions connect neighbouring Mn$^{2+}$ cations to give the Mn--O--C--O--Mn pathway that supports magnetic superexchange \cite{Inoue_1970,Scatena_2020,Wang_2004,Walker_2017}. Bulk thermodynamic measurements have shown the system to exhibit Curie-Weiss behaviour---thus no sign of magnetic ordering---for $T>2$\,K, with an antiferromagnetic coupling strength $ J\sim1$\,K \cite{Aston_2017}.

In this Letter, we show that Na[Mn(HCOO)$_3$] exhibits three hallmarks of geometric frustration: (i) the suppression of its N{\'e}el temperature $T_{\rm N}$ far below the characteristic energy scale of its magnetic interactions; (ii) the strongly-correlated disorder indicative of a CSL phase; and (iii) a magnetisation pseudo-plateau indicative of complex field-dependent behaviour. The magnetic ground-state is different to that expected for the trillium nnHAF model, but can be rationalised by accounting for the weaker but ever-present dipolar interactions (`nnHAF+D'). We show that the resulting nnHAF+D model: (i) accounts quantitatively for the magnetic susceptibility above $T_{\rm N}$ using an interaction strength $J$ that is consistent with $T_{\rm N}$; (ii) correctly predicts both the short-range magnetic correlations of the CSL phase and the magnetic ground-state, as observed in neutron-scattering measurements; and (iii) can be used to rationalise the existence and temperature-dependence of the measured magnetisation pseudo-plateau. A characteristic magnetocaloric effect associated with this pseudo-plateau emerges. 

\begin{figure}
  	\includegraphics[width=\columnwidth]{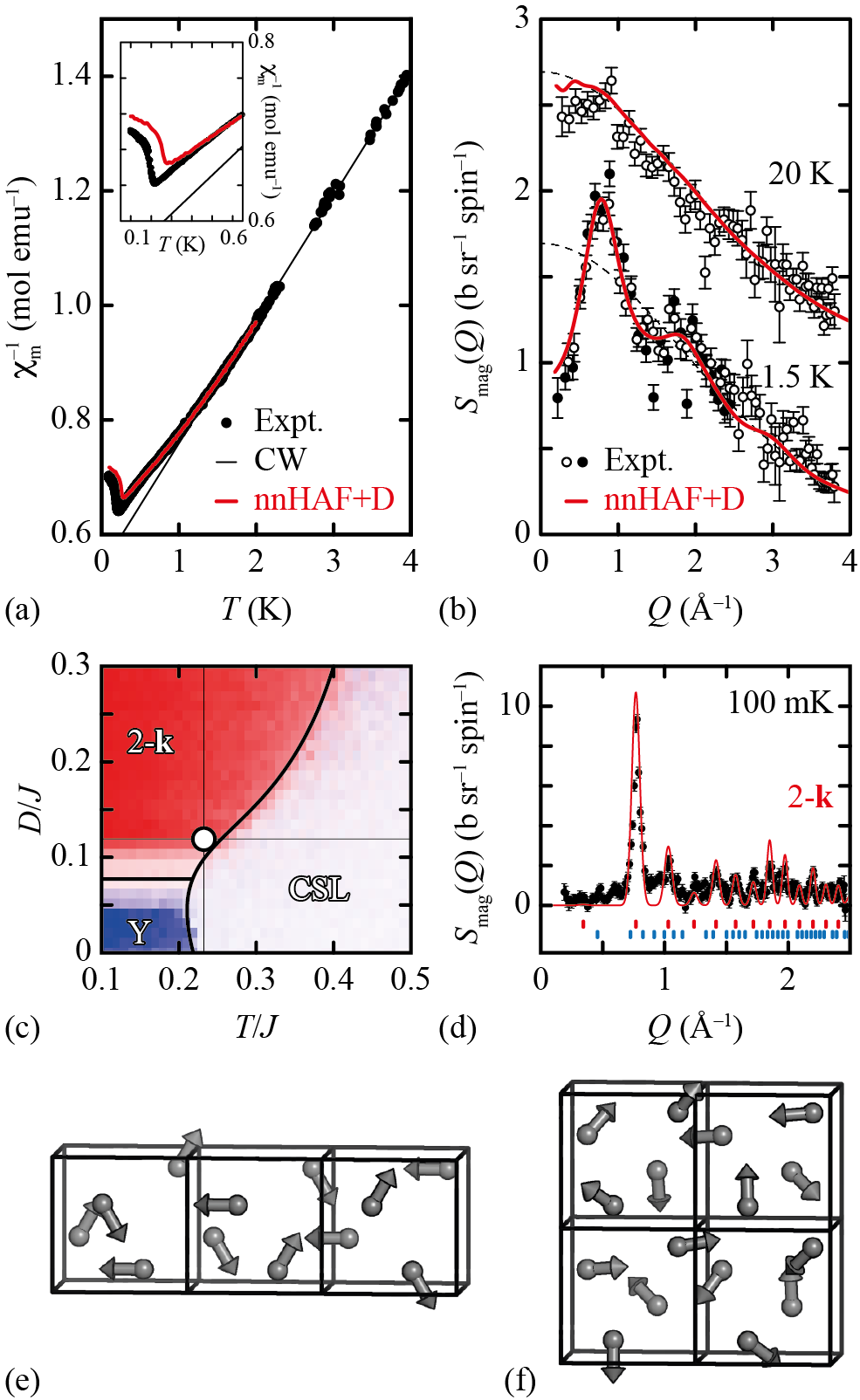}
  	\caption{\label{fig2} (a) Temperature dependence of the inverse magnetic susceptibility of Na[Mn(HCOO)$_3$]. Data shown as black solid circles; Curie-Weiss fit as a black line; and computed values for the nnHAF+D model as a red line. The inset highlights the region near $T_{\rm N}$. (b) The magnetic diffuse neutron scattering functions measured at 20 and 1.5\,K; open and filled circles denote data collected using neutron wavelengths of 3.1565 and 4.873\,\AA, respectively. The 20\,K values have been shifted vertically by one unit, and the Mn$^{2+}$ magnetic form factor is shown as a dashed line. (c) Magnetic ordering in the nnHAF+D model from MC simulations. The degree of red/blue colouring is taken from the strength of magnetic scattering associated with $\mathbf q_{\rm ord}=\langle\frac{1}{2},0,0\rangle$ and $\langle\frac{1}{3},0,0\rangle$, respectively. Experimental $T_{\rm N}$ and $D/J$ values are shown as vertical and horizontal lines. (d) The experimental magnetic neutron scattering function at 100\,mK (black markers) and that calculated directly from nnHAF+D simulations at $D/J=0.119$ (red line). Blue and red tick-marks denote the allowed magnetic reflection positions for the Y and 2-$\mathbf k$ phases, respectively. (e,f) Representations of the (e) Y and (f) 2-$\mathbf k$ magnetic structures.}
  \end{figure}

We begin by reporting the low-field magnetic behaviour of Na[Mn(HCOO)$_3$], expanding on the earlier measurements of Ref.~\citenum{Aston_2017}. We prepared and characterised a sample ($\sim 0.4$\,g) of Na[Mn(HCOO)$_3$] as described in \cite{si}. The low-field magnetic susceptibility, measured for a 10\,mg fraction of this sample, is shown as its inverse in Fig.~\ref{fig2}(a). For $T\gtrsim2$\,K, the magnetic behaviour is Curie-Weiss-like and antiferromagnetic; the corresponding fit over the range $2.0<T<4.5$\,K gives $\Theta_{\rm CW}=-2.3(3)$\,K, which is consistent with the value obtained in \cite{Aston_2017}. A N{\'e}el transition---seen here for the first time for this system---is observed at $T_{\rm N}=0.223(18)$\,K. The corresponding frustration parameter $f=|\Theta_{\rm CW}|/T_{\rm N}\sim10$ indicates strong frustration \cite{Savary_2016}.

Evidence for strong magnetic correlations within the CSL regime comes from the presence of structured magnetic diffuse scattering in polarised neutron scattering measurements. We used the D7 instrument at the ILL \cite{Stewart_2009,dois}, operating in XYZ polarisation mode, to separate the magnetic scattering for a powder sample of Na[Mn(HCOO)$_3$] from the nuclear and spin-incoherent components. Our measurements  at 20\,K ($T\gg J$) and 1.5\,K ($T\sim J$) are shown in Fig.~\ref{fig2}(b), where we have placed the scattering intensities on an absolute scale by Rietveld refinement of the nuclear scattering. At 20\,K, the magnetic scattering function $S_{\rm mag}(Q)$ is featureless and follows closely the Mn$^{2+}$ magnetic form factor \cite{Brown_2004}. By contrast, within the CSL regime at 1.5\,K we observe clear oscillations in $S_{\rm mag}(Q)$ characteristic of short-range magnetic order.

The magnetic scattering at 100\,mK ($T<T_{\rm N}\ll J$) reflects long-range magnetic order, but the magnetic Bragg diffraction pattern cannot be indexed in terms of the $\mathbf k_{\rm ord}= [\frac{1}{3},0,0]$ ordering vector expected from theory [Fig.~\ref{fig2}(d,e)] \cite{Hopkinson_2005,Isakov_2008}. Hence the physics governing selection of this magnetic ground-state must be more complex than accounted for by the simple nnHAF model of Refs.~\citenum{Hopkinson_2005,Isakov_2008}. 

An obvious omission of this model is consideration of the always-present magnetic dipolar interactions; note that no significant single-ion anisotropy is expected for the high-spin Mn$^{2+}$ ($d^5$) configuration in an octahedral crystal field. Using the crystallographic Mn$\ldots$Mn separation and taking $S=\frac{5}{2}$ for Mn$^{2+}$, the relevant dipolar energy scale in Na[Mn(HCOO)$_3$] is $D=0.118$\,K. Since $D\ll J$, one might ordinarily have expected the nnHAF ground-state to be robust to this additional interaction. Yet, in the absence of the corresponding theory (to the best of our knowledge), it fell to us to test this assertion by introducing dipolar interactions into the nnHAF model.

In this spirit, we carried out a series of classical Monte Carlo (MC) simulations related to those of Ref.~\citenum{Isakov_2008} but with MC energies now calculated using the nnHAF+D expression
\begin{equation}\label{hamil}
    E_{\rm MC}=J\sum_{\langle i>j\rangle}\mathbf S_i\cdot\mathbf S_j +D\sum_{i>j}\frac{\mathbf S_i\cdot\mathbf S_j+3(\mathbf S_i\cdot\hat{\mathbf r}_{ij})(\mathbf S_j\cdot\hat{\mathbf r}_{ij})}{(r_{ij}/r_1)^3}.
\end{equation}
Here, $\mathbf S_i$ represents the vector spin orientation at site $i$, $\hat{\mathbf r}_{ij}$ is the unit-vector pointing from spin $i$ to spin $j$, and $r_1$ is the nearest-neighbour separation. The first sum is taken over distinct nearest-neighbour pairs $i,j$. We used Ewald summation as implemented in  Ref.~\citenum{Paddison_2015} to treat the long-range dipolar interactions. Here, as elsewhere, we subsume the spin magnitude within the constants $J$ and $D$. Our MC configurations respresented $6\times6\times6$ supercells of the $P2_13$ unit-cell shown in Fig.~\ref{fig1}(a); full details of these simulations and our methods for extracting physical observables are given in \cite{si}.

The type of magnetic order driven by Eq.~\eqref{hamil} at low temperatures turns out to be surprisingly sensitive to the value of $D/J$ [Fig.~\ref{fig2}(c)]. For $D\gtrsim 0.08J$, the magnetic ground-state switches from the helical $\mathbf k_{\rm ord}= [\frac{1}{3},0,0]$ order of the nnHAF model to a 2-$\mathbf k$ state with $\mathbf k_{\rm ord}\in \langle\frac{1}{2},0,0\rangle$ [Fig.~\ref{fig2}(f)]. This sensitivity is evident also from mean--field calculations (see \cite{si}). The magnetic Bragg diffraction pattern of this new 2-$\mathbf k$ ground-state shows a remarkable similarity to our 100\,mK measurement for Na[Mn(HCOO)$_3$] [Fig.~\ref{fig2}(d)], which suggests that the nnHAF+D model may indeed capture the key physics at play in this magnetic MOF.

In order to test the effectiveness of the model against all our experimental measurements, we required an accurate estimate of the Heisenberg exchange strength $J$. We found that for $D\ll J$, the behaviour of Eq.~\eqref{hamil} at $T\geq J/2$ was essentially indistinguishable to the nnHAF model ($D=0$; see \cite{si}). Consequently, we estimated $J$ by determining the $T\geq J/2$ susceptibility for the $D=0$ case, and then carrying out a least-squares fit of $J$ to best match experiment (see \cite{si}). In this way we obtain $J=0.96(2)$\,K, which is consistent with the estimate given in Ref.~\citenum{Aston_2017}. Moreover, the resulting relative energy scale $D/J=0.119$ locates Na[Mn(HCOO)$_3$]  within the regime for which 2-$\mathbf k$ order is expected [Fig.~\ref{fig2}(c)].

The results of our nnHAF+D MC simulations carried out using this combination of $J$ and $D$ are shown as the red solid lines in Fig.~\ref{fig2}(a) and (b). Considering first the magnetic susceptibility data, our simulations capture at once the Curie-Weiss behaviour for $T\gtrsim2$\,K, the departure from Curie-Weiss behaviour for $T\lesssim2$\,K, and the N{\'e}el transition itself. We find $T_{\rm N}=0.27(1)$\,K, in good agreement with experiment. Likewise for our magnetic neutron scattering data: the $S_{\rm mag}(Q)$ functions calculated directly from MC configurations generated at the corresponding temperatures \cite{Blech_1964} follow experiment closely (see \cite{si} for the corresponding spin correlation functions). And, of course, the MC ground-state is a 2-$\mathbf k$ structure for which $S_{\rm mag}(Q)$ is consistent with the 100\,mK neutron magnetic diffraction pattern [Fig.~\ref{fig2}(d)].

So the nnHAF+D Hamiltonian provides a good representation of the spin physics at play in Na[Mn(HCOO)$_3$]. Perhaps the largest discrepancy between experiment and calculation is the small difference in $T_{\rm N}$ [see inset to Fig.~\ref{fig2}(a)]. This discrepancy may reflect the relevance of yet weaker interactions, such as single-ion anisotropy (usually safely ignored for high-spin Mn$^{2+}$) and/or antisymmetric exchange interactions, as allowed by the chiral crystal structure. We have not explored either case here further. We do note, however, that the dipolar interactions of our nnHAF+D model induce an Ising-like spin anisotropy consistent with the crystallographic point symmetry of the Mn site (\emph{cf}.~\cite{Paddison_2015}; see \cite{si}). Consequently, an easy-plane single-ion term in an extended Hamiltonian may compete with the dipolar interactions, in turn reducing $T_{\rm N}$.

\begin{figure}
	\includegraphics{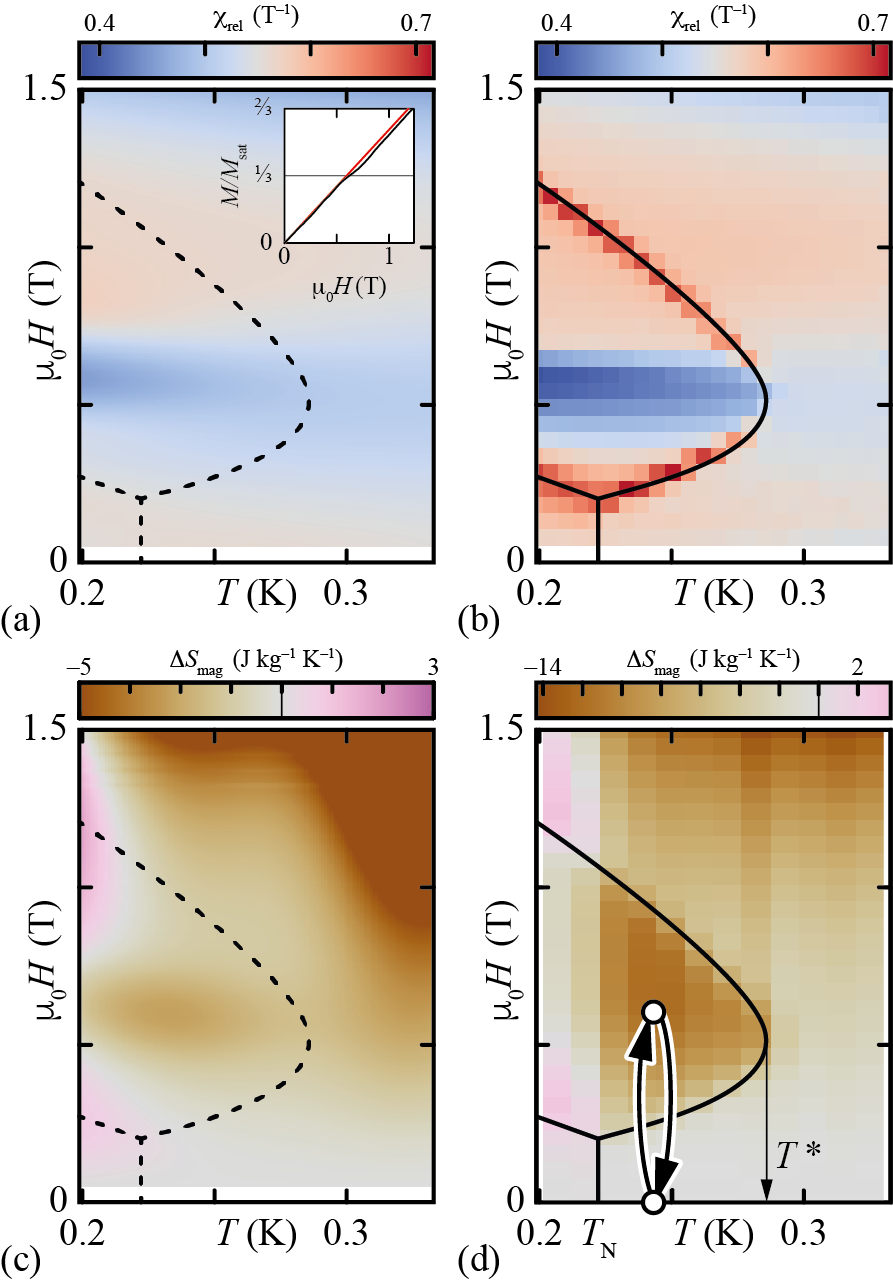}
	\caption{\label{fig3} (a) Experimental field and temperature dependence of the magnetic susceptibility of Na[Mn(HCOO)$_3$]. The inset shows the field-dependent magnetisation at 100\,mK; experimental data in black and low-field linear fit in red. Note the pseudo-plateau at $M/M_{\rm sat}\simeq\frac{1}{3}$. This feature corresponds to the horizontal blue region at $\sim$0.6 T in the main plot. (b) Magnetic susceptibility from MC simulations, with phase boundaries shown as solid lines. (c,d) The corresponding field-induced change in magnetic entropy from (c) experiment and (d) MC simulations, which has a local maximum within the UUD phase. Within the regime $T_{\rm N}<T<T^\ast$, it is possible to cycle between disordered and ordered states under the application and removal of a low magnetic field. Hence the cycling regime illustrated by white circles in panel (d) offers a magnetocaloric cooling strategy.}
\end{figure}

Given the sensitivity of the nnHAF model to perturbations in the zero-field limit, we might expect Na[Mn(HCOO)$_3$] to be especially sensitive to an applied magnetic field. To explore this point we carried out a series of high-field magnetisation measurements at finely-spaced temperatures within the ordered and CSL regimes ($200<T<350$\,mK) using applied fields $0<\mu_0 H<1.5$\,T. Our results are summarised in Fig.~\ref{fig3}(a), where we plot the susceptibility $\chi_{\rm m}={\rm d}(M/M_{\rm sat})/{\rm d}\mu_0H$ as a function of $\mu_0H$ and $T$. The most obvious feature we observe is a valley in $\chi_{\rm m}$ centred on $\mu_0H\sim0.6$\,T that corresponds to a pseudo-plateau in the  field-dependence of the magnetisation at $M/M_{\rm sat}=\frac{1}{3}$ [see inset to Fig.~\ref{fig3}(a)]. This feature is strongest at low temperatures, but persists with varying prominence throughout the temperature range of our measurements. We proceed to show that the origin of this feature is the suppression of magnetic fluctuations within a new ordered phase that is stabilised at finite field.

To help understand this field-dependant behaviour, we explored the effect of introducing a Zeeman term into the original nnHAF model:
\begin{equation}\label{field}
    E=J\sum_{\langle i>j\rangle}\mathbf S_i\cdot\mathbf S_j+\mu_0\mathbf H\cdot\sum_i\mathbf S_i.
\end{equation}
Here we are intentionally neglecting dipolar interactions for two key reasons: (i) the corresponding energy scale in this system is an order of magnitude smaller than that of Heisenberg exchange, and (ii) our powder measurements represent an average over crystallite orientations, the effect of which is computationally difficult to take into account when including anisotropic interactions. We will discuss in due course the most important consequences of this approximation. Nevertheless we carried out a new set of MC simulations driven by the modified Hamiltonian \eqref{field}, focusing on the same range of $\mu_0H$ and $T$ values explored experimentally. 

The simulation results are shown in Fig.~\ref{fig3}(b) and reflect both similarities and differences to experiment. Perhaps most importantly, the susceptibility minimum near $M/M_{\rm sat}\sim\frac{1}{3}$ appears strongly for temperatures below 290\,mK, which indicates that a magnetisation pseudo-plateau is intrinsic to the Hamiltonian \eqref{field}. In the simulations, this susceptibility `valley' is flanked by two sharp maxima that demarcate a new phase region centred at finite field. This phase is stable for temperatures beyond $T_{\rm N}$ to a critical value $T^\ast\sim290$\,mK. The susceptibility maxima marking the phase boundary are not resolved in our experimental data, but may have been washed out as a result of orientational averaging. The zero-field nnHAF Hamiltonian \eqref{field} is isotropic, which means the system's response is invariant to the direction of applied field and hence there is no need to take into account crystallite orientation; the presence of anisotropic dipolar interactions in practice introduces an orientational dependence over which our experimental measurements are integrated.

Inspection of our MC simulations shows the new field-stabilised phase to be an `up--up--down' (UUD) state in which the moments of two thirds of the Mn$^{2+}$ ions align parallel to the applied field, and that of the other third opposes the field. Complete UUD order gives $M/M_{\rm sat}=\frac{1}{3}$, which is why the susceptibility drops around this value. We note that the field-dependent behaviour of Heisenberg antiferromagnets on other tripartite lattices (\emph{e.g.}\ triangular \cite{Kawamura_1985,Seabra_2011} and Shastry-Sutherland (SSL) \cite{Moliner_2009}) are also known to involve a transition from their zero-field ground-state to an `UUD' phase, such as we see here \cite{Kawamura_1985,Seabra_2011,Moliner_2009}. The similarity to SSL physics turns out to be a geometric feature of the trillium net; a discussion of the mapping between the two systems is given in \cite{si}.

There is a special significance of the proximity of (disordered) CSL and (ordered) UUD phases for $T_{\rm N}<T<T^\ast$. Within this regime, an applied field drives a disorder/order transition, resulting in a sudden loss of entropy to be expelled as heat---\emph{i.e.} a magnetocaloric effect \cite{Kitanovski_2015}. Moreover, since the UUD state has $M/M_{\rm sat}\sim\frac{1}{3}$, the field required to generate this entropy change is small by comparison to that required to induce transitions in conventional antiferromagnets. As such we expect an attractive low-field magnetocaloric entropy change in Na[Mn(HCOO)$_3$].

The magnitude of this magnetocaloric response was calculated from both experiment and simulation according to the relation \cite{Kitanovski_2015}
\begin{equation}
\Delta S_{\rm mag}(H,T) = \mu_0 \int_0^H \frac{\partial M_{\parallel}(H^\prime,T)}{\partial T} {\rm d}H'.
\end{equation}
Here $M_{\parallel}$ is the magnetisation parallel to the direction of applied field. Our results are shown in Fig.~\ref{fig3}(c,d). For temperatures $T>T^\ast$, the magnetic entropy change behaves essentially as one expects for a paramagnet: field increasingly reduces the magnetic degrees of freedom, such that entropy reduces smoothly and monotonically. For $T<T^\ast$, however, there is a local maximum in magnetic entropy change that coincides with UUD order. In our simulations, the most negative value of $\Delta S_{\rm mag}$ we obtain (\emph{viz.}\ $-12$\,J\,kg$^{-1}$\,K$^{-1}$) is actually competitive with that of systems used commercially for magnetocaloric cooling \cite{Numazawa_2003,Mukherjee_2017}. The experimental value is less extreme by a factor of about four because orientational averaging over crystallite orientations makes the CSL$\rightarrow$UUD transition smooth. One expects a sharper transition in single-crystal measurements that no longer integrate over crystallite orientation, which in turn would amplify the magnetocaloric entropy change attainable in this system. One way or the other, the basic principle would be to cycle between CSL and UUD states by applying and removing field, dumping the heat generated as entropy during the CSL$\rightarrow$UUD transition, then forcing the system to cool its environment on recovery of the higher-entropy CSL phase [Fig.~\ref{fig3}(d)].
 
From a materials design perspective, avenues for optimising trillium magnetocalorics are straightforwardly identified---after all, the phase behaviour we identify here depends only on the magnitudes of $S$ and $J$, and the trillium edge length, which in turn influences $D$. Because a large $D/J$ washes out the magnetocaloric effect in powder samples, optimising the isotropic coupling $J$ is likely better than optimising $S$. One possibility for doing so is to use an alternative bridging ion (\emph{e.g.}\ azide) with stronger superexchange.

We conclude by highlighting the importance of geometric frustration within the CSL phase for generating a strong magnetocaloric effect at relatively low applied fields. A number of the most important magnetocaloric materials, such as gadolinium gallium garnet, are also strongly frustrated \cite{Saines_2015, Schiffer_1994,Zhitomirsky_2003,Schmidt_2007}, and we see our analysis here as providing a robust link between frustration and this useful physical property. A corollary is that non-magnetic frustration in pseudospin systems may favour strong barocaloric responses at low pressures, offering a new design strategy for efficient solid-state cooling devices \cite{Lloveras_2019, Aprea_2020,Simonov_2020}.
 
\begin{acknowledgments}
The authors gratefully acknowledge financial support from the E.R.C. (Grant 788144), E.P.S.R.C. (Grant EP/T027886/1) and the Leverhulme Trust (Grant RPG-2018-268). JAMP's work at Cambridge was supported by Churchill College, Cambridge. JAMP's work at ORNL was supported by the Laboratory Directed Research and Development Program of Oak Ridge National Laboratory, managed by UT--Battelle, LLC for the U.S. Department of Energy. JMB and ALG gratefully acknowledge assistance from Philip Welch (Oxford). Low temperature magnetization experiments were funded by the European Union’s Horizon 2020 Research and Innovation Programme, under Grant Agreement no 824109.
\end{acknowledgments}

\end{document}